\documentclass[%
reprint,
superscriptaddress,
preprint,
bibnotes,
amsmath,amssymb,
aps,
reprint,
prb,
]{revtex4-2}

\usepackage{graphicx}% Include figure files
\usepackage{dcolumn}% Align table columns on decimal point
\usepackage{bm}% bold math
\usepackage{physics}
\usepackage{siunitx}
\usepackage[normalem]{ulem}
\usepackage{xprintlen}
\usepackage[T1]{fontenc}
\usepackage[utf8]{inputenc}
\usepackage{romannum}
\usepackage{natbib}
\usepackage{layouts}
\usepackage{float}
\usepackage{amsmath}

\usepackage[caption=false]{subfig}

\usepackage{todonotes}

\usepackage[colorlinks=true, citecolor=blue]{hyperref}
\usepackage[capitalize]{cleveref}   % ,noabbrev

\usepackage{color}
\usepackage{colordvi} 
\definecolor{Red}{rgb}{1,0,0}

\definecolor{blue}{rgb}{0,0,1}

\DeclareUnicodeCharacter{0119}{\k{e}}

\begin{document}
%\the\columnwidth
\title{Coherent control and high purity single-photon emission from telecom quantum dots on metamorphic InGaAs}
\author{Beatrice Costa}
\email{beatrice.costa@tum.de}
\thanks{ORCID: 0009-0007-9528-6206}
\affiliation{Department of Electrical and Computer Engineering, TUM School of Computation, Information and Technology, Technical University of Munich}
\affiliation{Munich Center for Quantum Science and Technology (MCQST)}
\author{Bianca Scaparra}
\affiliation{Walter Schottky Institute, TUM School of Natural Sciences, Technical University of Munich}
\author{Hubert Riedl}
\affiliation{Walter Schottky Institute, TUM School of Natural Sciences, Technical University of Munich}
\author{Gregor Koblmüller}
\affiliation{Munich Center for Quantum Science and Technology (MCQST)}
\affiliation{Walter Schottky Institute, TUM School of Natural Sciences, Technical University of Munich}
\affiliation{Institute of Physics and Astronomy, Technical University Berlin}
\author{Eugenio Zallo}
\affiliation{Munich Center for Quantum Science and Technology (MCQST)}
\affiliation{Walter Schottky Institute, TUM School of Natural Sciences, Technical University of Munich}
\author{Jonathan Finley}
\affiliation{Munich Center for Quantum Science and Technology (MCQST)}
\affiliation{Walter Schottky Institute, TUM School of Natural Sciences, Technical University of Munich}
\author{Lukas Hanschke}
\affiliation{Department of Electrical and Computer Engineering, TUM School of Computation, Information and Technology, Technical University of Munich}
\affiliation{Munich Center for Quantum Science and Technology (MCQST)}
\author{Kai Müller}
\affiliation{Department of Electrical and Computer Engineering, TUM School of Computation, Information and Technology, Technical University of Munich}
\affiliation{Munich Center for Quantum Science and Technology (MCQST)}
\date{\today}

\begin{abstract}
Semiconductor quantum dots emitting in the telecommunication O- and C-bands are promising candidates as single-photon sources for fiber-based quantum communication. In this work, we show how coherent quasi‑resonant and phonon‑assisted excitation help reveal the intrinsic optical properties of InAs quantum dots embedded in relaxed InGaAs matrices on top of metamorphic InGaAs layers for achieving emission in the telecom bands. By systematically comparing excitation schemes, we separate excitation‑induced effects from intrinsic emitter properties. This allows us to observe linewidths as narrow as \SI{21}{\micro eV}, the disappearance of slow decay channels, and single‑photon purity reaching second-order autocorrelation values in the order of $10^{-2}$. Photoluminescence excitation reveals higher excited states, enabling coherent control of trion transitions and the observation of Rabi‑like oscillations in both telecom bands. These results confirm the promising potential of these quantum dots for applications in quantum technologies.
\end{abstract}

\maketitle

Semiconductor quantum dots (QDs) emitting in the near-infrared and telecommunication ranges are promising candidates for on‑demand single‑photon sources, due to their excellent optical properties \cite{senellart2017high, hanschke2018quantum, Schoell2019RF, basso2021quantum, sbresny2022stimulated}. In particular, QDs emitting in the telecommunication O- and C-bands are being investigated for their potential as sources of single photons and entangled photon pairs compatible with long distance fiber-based quantum communication infrastructure \cite{muller2018quantum, jeon2022plug, vajner2022quantum, gyger2022metropolitan, Vajner2024ondemand, Holewa2025telecom}. The two principal material systems to achieve quantum dot emission in the telecommunication bands are indium arsenide (InAs) QDs on indium phosphide (InP) and InAs QDs on gallium arsenide (GaAs) substrates using strain-engineered InGaAs layers. Such QDs have recently shown promising progress towards transform-limited emission linewidths and improved single‑photon purity under quasi-resonant and resonant excitation schemes \cite{Nawrath2019Coherence, Nawrath2021RF, AlKhuzheyri2016RF, Wells2023Rabi, phillips2024purcell}, bringing their performance closer to the stringent requirements of quantum communication protocols. 

The structure examined in this work consists of InAs telecom QDs embedded in mostly relaxed InGaAs matrices grown on graded InGaAs buffers and integrated with nearly lattice-matched DBRs grown by molecular beam epitaxy (MBE). Their growth is described in detail in Ref. \cite{Scaparra2024broad}. So far, these QDs have only been characterized using above-bandgap excitation \cite{Costa2026telecom}. However, under nonresonant excitation, the intrinsic optical properties of the emitters are obscured due to the introduction of excess carriers, charge noise, and recapture processes. Therefore, the intrinsic optical properties of these emitters, particularly their linewidth, radiative lifetime, and single-photon purity, remain largely unexplored.

Importantly, it has previously been demonstrated on different platforms that quasi-resonant excitation schemes, either addressing higher excited states of the QDs or their longitudinal-acoustic (LA)-phonon sidebands, suppress the generation of free carriers and wetting layer recapture \cite{dalgarno2008hole, KuhlmannTransform2015}. This yields significant improvements not only on the observed linewidths (which approach transform limit for well-engineered systems \cite{Albrechtsen2025Quantum}), but also in their single-photon purity with the second order autocorrelation $g^{(2)}(0)$ values in the order of $10^{-2}$ \cite{Dusanowski2017purity, Zeuner2021entangled, Joos2024SUPER, Hauser2026determ, Vyvlecka2023LA}). Furthermore, excitation via higher excited states enables coherent control of the emitter \cite{Wells2023Rabi},  whereas LA-phonon assisted excitation preserves photon indistinguishability and enables the generation of cluster states and polarization-encoded spin-multiphoton entanglement \cite{ Lindner2009LA, Cosacchi2019LA, Thomas2021LA, Coste2023Phonon,  Lehner2023LA}.

In the present work, we explicitly compare four excitation schemes: above-bandgap (ABE), wetting layer (WLE), coherent quasi-resonant (QRE), and LA phonon-assisted (PAE), across QDs emitting in both the O- and C-bands. By separating the role of the excitation scheme from the intrinsic properties of the emitter, we reveal that the optical quality of these QDs is significantly better than previously reported measurements suggested. In particular, we observe narrow linewidths down to \SI{21}{\micro eV} and $g^{(2)}(0)$ values on the order of $10^{-2}$. By identifying resonances corresponding to excited states of the QDs we furthermore gain insight into their energetic structure and achieve coherent control of these states. These findings establish the intrinsic optical quality of the metamorphic GaAs-based QD platform and its suitability for applications.

In particular, two different samples are investigated, where the emission is tailored to the O-band (sample O) and to the C-band (sample C). The optical characterization of single QDs was carried out using a conventional confocal microscopy $\mu$-PL setup. The samples were placed in a closed-cycle cryostat operating at \SI{4}{ K}. The excitation was provided by a non-resonant continuous-wave (CW) laser at \SI{898}{ nm} and by a tunable pulsed Ti:Sapphire laser with a repetition rate of \SI{80}{ MHz} and tuned to the desired wavelength by an OPO. The laser pulses, originally in the femtosecond range, were stretched to picosecond timescales and centered at the desired excitation wavelength using a custom-built 4f pulse shaper. The PL signal was analyzed using a grating spectrometer and detected by an InGaAs CCD. The gratings used to record the spectra for the two samples have a resolution limit of \SI{20}{\micro eV} at \SI{1310}{nm} and \SI{25}{\micro eV} at \SI{1550}{nm}, respectively. 

First, we examine the linewidth of the QDs to determine how it is influenced by the choice of excitation scheme. For this, we chose to perform all measurements on trion (X*) transitions, due to their higher brightness and lack of fine-structure splitting. 
The excitation schemes and wavelengths used are above-bandgap excitation at \SI{900}{nm}, wetting layer excitation at \SIrange{1200}{1220}{nm} for sample O and \SIrange{1300}{1350}{nm} for sample C, quasi-resonant excitation addressing resonances corresponding to higher excited states of the QDs, and phonon-assisted addressing the LA-phonon sidebands of the QDs at \SIrange{5}{0.5}{meV} detuning.

\begin{figure}[h!]
\centering
\includegraphics[width=0.5\textwidth]{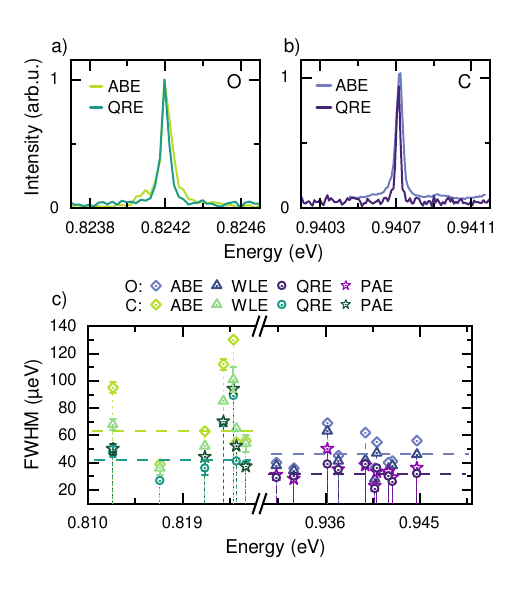}
\caption{ a) Spectra of the same trion (X*) on sample O, observed under ABE and QRE, showing the narrowing of the FWHM to resolution-limited values and disappearance of background. b)  Spectra of an X* transition on sample C, observed under the same excitation techniques. c) X* linewidths (FWHM) of samples O (purple) and C (green), under ABE, WLE, QRE, and PAE excitations. For sample O, the median FWHM of \SI{43 \pm 1}{\micro eV} under ABE decreases to \SI{31 \pm 1}{\micro eV} under QRE, indicated by dashed lines on the plot. Sample C shows a similar reduction in FWHM from \SI{63 \pm 2}{\micro eV} to \SI{41 \pm 2}{\micro eV}.} 
\label{FWHM}
\end{figure}

Figures \ref{FWHM}a) and b) show representative spectra of trion (X*) transitions under above-bandgap (lighter color) and quasi-resonant excitation (darker color, addressing a higher excited state of the QD), for sample O and sample C, respectively. For the studied transition on sample O, shown in Fig. \ref{FWHM}a), the observed FWHM value shows an improvement from \SI{36 \pm 1}{\micro eV} under above-bandgap excitation to \SI{21 \pm 1}{\micro eV} under quasi-resonant excitation. For the one on sample C, shown in Fig. \ref{FWHM}b), it decreases from \SI{55 \pm 1}{\micro eV} (above-bandgap) to \SI{39 \pm 1}{\micro eV} (quasi-resonant). In both cases, a significant reduction of the emission linewidth can be observed, with some of the observed linewidths narrowing down to the spectrometer resolution.

For a statistical analysis, the X* transitions of $17$ QDs were studied in each sample under the four excitation techniques. The emission spectra were fitted with Lorentzian and Gaussian functions to estimate the FWHM.
Figure \ref{FWHM}c) shows the extracted linewidth values. It shows a decrease from a median value of \SI{43 \pm 1}{\micro eV} for sample O under above-bandgap excitation (light purple dashed line) to a median value of \SI{31 \pm 1}{\micro eV} under quasi-resonant excitation (dark purple dashed line). For sample C, the median FWHM narrows from \SI{63 \pm 2}{\micro eV} under above-bandgap excitation (light green dashed line) to \SI{41 \pm 2}{\micro eV} under quasi-resonant excitation (dark green dashed line).
Therefore, about $30\%$ narrowing can be seen from above-bandgap to quasi-resonant excitation on both samples, with some of the linewidths obtained under quasi-resonant excitation reaching the resolution limit of the spectrometer gratings used for the measurements (\SI{25}{\micro eV} for sample C and \SI{20}{\micro eV} for sample O).

Previously reported values of linewidths of X* for MBE-grown InAs QDs on InP substrates, obtained under quasi-resonant excitation, include \SIrange{29}{34}{\micro eV} \cite{Kim2025TPI}, and \SI{460}{\micro eV} on GaAs substrates \cite{Wyborski2023Impact}. 
For InAs QDs grown via MOVPE on GaAs substrate, a value of \SI{65 \pm 2}{\micro eV} was reported \cite{Nawrath2023Bright}.

The resonances used for quasi-resonant excitation were identified using photoluminescence excitation (PLE) measurements. The scans were conducted by scanning the excitation wavelength of the pulsed laser in a wavelength range spanning the end of the wetting layer to approximately \SI{0.5}{meV} detuning from the s-shell of the examined trion. The resulting emission was spectrally filtered using a filter similar to the 4f pulse shaper and detected with superconducting nanowire single-photon detectors. Time-tagging of detected photons was implemented using time-correlated single-photon counting module.

\begin{figure}[h!]
\centering
\includegraphics[width=0.5\textwidth]{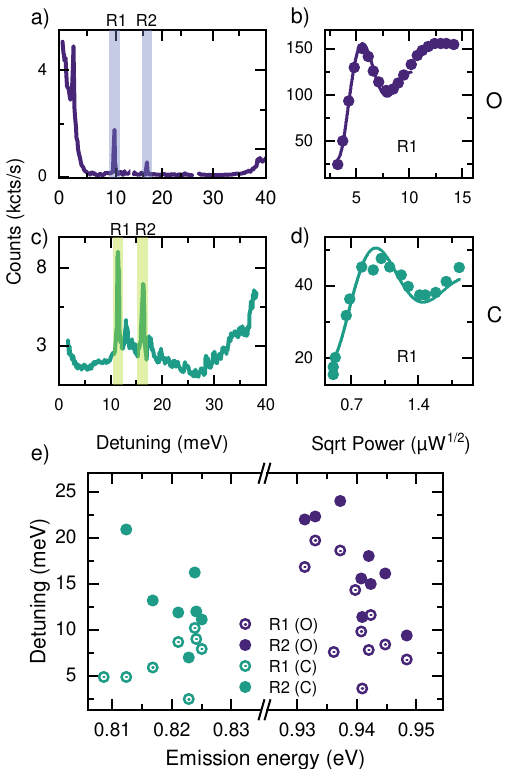}
\caption{a) PLE scan spanning from the LA-phonon sideband to the end of the wetting layer, performed at \SI{1}{\micro W} excitation power on a trion (X*) on sample O. The resonances highlighted in green can be attributed to higher excited states of the QD, at \SI{9.8}{meV} (R1) and \SI{15.5}{meV} (R2) detuning. b) Evidence of Rabi-like oscillatory behaviour on the same X* in sample O, under QRE addressing R1. c) PLE scan performed under comparable excitation conditions on a X* on sample C, with resonances (highlighted in purple) at \SI{10.2}{meV} (R1) and \SI{16.2}{meV} (R2) detuning. d) Evidence of Rabi-like oscillatory behaviour on the same X* in sample C. e) Statistics of the detuning energies of resonances R1 and R2 observed from PLE scans conducted on $19$ QDs.} 
\label{PLE}
\end{figure}

Figures \ref{PLE}a) and c) show PLE scans conducted on X* transitions on samples O and C, respectively. In both cases, we can find the tail of the wetting layer (WL) at an approximate detuning of \SIrange{40}{30}{meV} from the s-shell, two main isolated resonances and, from approximately \SI{5}{meV} detuning, the start of the LA-phonon sideband. Both scans were performed at a power of \SI{1}{\micro W} and with approximate pulse durations of \SI{10}{ps} and \SI{16}{ps}, respectively. For clarity, we label the observed resonances R1 and R2 according to their energetic proximity to the s-shell transition, with R1 denoting the closest resonance. The resonances highlighted in Fig. \ref{PLE}a) correspond to detunings of the examined transition from the s-shell of \SI{9.8}{meV} for R1 and \SI{15.5}{meV} for R2. The resonances highlighted in Fig. \ref{PLE}c) correspond to a detuning of \SI{10.2}{meV} for R1 and R2 at \SI{16.2}{meV} respectively.

Figures \ref{PLE}b) and d) show the power-dependent emission intensity obtained by exciting the R1 resonance for sample O and for sample C. In both samples, the emission intensity exhibits clear oscillations characteristic of Rabi oscillations, demonstrating coherent optical control of QDs emitting in both telecom wavelength bands. It can be seen on both samples that the oscillations are damped and that it is only possible to observe Rabi oscillations up to pulse areas of $3\pi$. Damping of Rabi oscillations is likely associated with rapid loss of coherence arising from mechanisms such as motional narrowing, charge-carrier tunnelling \cite{Cassabois2008dephasing, Ramsay2010damping, Ardelt2015tunneling}, or the influence of WL states \cite{Villas2005decoherence, LöblExcitons2019}. However, the fact that the Rabi oscillations are damped towards the top points towards a short lifetime of the driven states compared to the pulse length \cite{Ardelt2014dissipative}, as will be analyzed in time-resolved measurements below.

Fig \ref{PLE}e) shows the statistics from PLE scans of $19$ QDs distributed between the two samples. On sample O, we observed R1 at a median detuning of \SI{9.8}{meV} (with standard deviation \SI{5.3}{meV} and range \SI{16.1}{meV}), and R2 at a median detuning of \SI{16.1}{meV} (with standard deviation \SI{5}{meV} and range \SI{14.6}{meV}). On sample C, we observed median detuning of \SI{6.9}{meV} (with standard deviation \SI{2.6}{meV} and range \SI{7.7}{meV}) for R1, and R2 at a median detuning of \SI{12}{meV} (with standard deviation \SI{4.4}{meV} and range \SI{13.9}{meV}). 
Despite the difference in indium content between the two samples, R1 and R2 appear within a restricted detuning range across all studied QDs, well below the onset of the wetting layer continuum at \SIrange{30}{40}{meV} detuning. This confirms that higher excited states can be reliably addressed on this platform while minimizing excitation of the wetting-layer continuum \cite{Goldmann2014structural,Podemski2021detuning, Kosarev2022model, Scaparra2024broad}. 

Such resonances at comparable detuning energies have been observed across a variety of material systems and growth techniques, suggesting that excited state detunings in such a range from the s-shell are consistent with other InAs-based telecom QD systems. The precise values do however vary between different platforms, as excited state energies depend significantly on material composition, strain, confinement geometry, and the charge configuration of the emitters \cite{Podemski2021detuning, Gao2022QKD, Wyborski2023Impact, Kim2025TPI, Hauser2026determ}.

While PLE scans reveal the internal energy level structure of the emitters, time-resolved measurements provide crucial information about their population dynamics. Comparing the decay dynamics under the four excitation schemes allows us to separate excitation-induced processes from the intrinsic s-shell X* lifetime. For instance, coherent quasi-resonant excitation enables the extraction of the effective relaxation time of the addressed excited state, linking the excited states identified in the PLE scans to the population transfer that feeds the emitting trion state.

\begin{figure}[h!]
\centering
\includegraphics[width=0.5\textwidth]{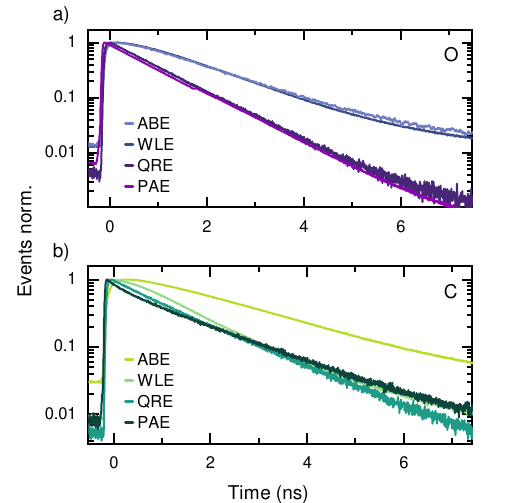}
\caption{a) Radiative decay of X* transition from sample O under the four excitation schemes. b) Comparable measurements obtained on a X* transition on sample C.}
\label{TI}
\end{figure}

Figure \ref{TI}a) shows the time-resolved PL of an X* transition from sample O under the four excitation schemes. Under above-bandgap excitation, the decay exhibits a pronounced delayed onset followed by a slower tail, with similar behavior observed under wetting-layer excitation. Under quasi-resonant and LA-phonon-assisted excitation, the slow tail disappears, while the delayed onset is strongly reduced, particularly for LA-phonon-assisted excitation. Figure \ref{TI}b) shows the corresponding measurements for sample C. In contrast to sample O, a slower component remains visible under LA-phonon-assisted excitation, resulting in a biexponential decay. This additional contribution may originate from carrier dynamics in the surrounding matrix, and it is consistent with the introduction of secondary slower population channels. 

Analysis of the quasi-resonant decay dynamics also reveals the relaxation time for the excited state addressed by the laser pulse. For the transitions shown in Figs. \ref{TI}a) and \ref{TI}b), we obtain \SI{28 \pm 0.2}{ps} and \SI{25 \pm 0.3}{ps}, respectively. These times are comparable to the corresponding excitation-pulse durations of \SI{10}{ps} and \SI{16}{ps}. Consequently, population relaxation occurs already during the pulse, limiting the population inversion that can be achieved. Together with intrinsic decoherence mechanisms, including coupling to LA-phonons, this contributes to the rapid damping of the Rabi oscillations observed in Fig. \ref{PLE}b) and d) \cite{Stufler2006Rabi, Calcagno2025Dynamically}. Across all examined dots, the effective excited-state relaxation times have median values of \SI{40}{ps} for sample O, spanning \SIrange{13}{171}{ps}, and \SI{57}{ps} for sample C, over a range of \SIrange{4}{126}{ps}, with no clear dependence on detuning or emission energy.

For comparison, measurements under above-bandgap excitation on the same samples show biexponential decay dynamics for most of the investigated excitonic complexes, with a fast decay component followed by a slower tail \cite{Costa2026telecom}. The presence and magnitude of these contributions depend strongly on the excitation scheme, as does the onset of the emission \cite{dalgarno2005dark, dangel2022two}. In particular, the delayed onset observed under above-bandgap and wetting layer excitation is consistent with carrier relaxation and recapture from the wetting layer \cite{Syperek2013Impact, Reithmaier2014Carrier}. These contributions are progressively suppressed under quasi-resonant and LA-phonon-assisted excitation due to the shorter lifetime of the excited state and the direct coupling to the s-shell, respectively \cite{Suzuki2018Dephasing, Lehner2023LA}. The resulting near-instantaneous decay start and reduced contribution from secondary channels enable a more reliable determination of the intrinsic X* lifetime, corresponding to \SI{0.95 \pm 0.001}{ns} for sample O in Fig. \ref{TI}a) and \SI{1.62 \pm 0.006}{ns} for sample C in Fig. \ref{TI}b).

Having established the influence of the excitation scheme on the linewidth and radiative dynamics, we next investigate its impact on the single-photon emission characteristics through second-order autocorrelation measurements. These measurements employed the same spectral filtering, single-photon detectors, and excitation laser described above, together with a Hanbury Brown and Twiss (HBT) interferometer consisting of a fiber-coupled beam splitter.

Figure \ref{g2}a) shows a second-order autocorrelation measurement performed on an X* transition of sample O, under above-bandgap excitation, conducted at a power of \SI{10}{\micro W}, corresponding to $21\%$ of saturation power, and with a pulse duration of \SI{6}{ps}. The measurement yields a $g^{(2)}(0) \text{ value of } 0.08 \pm 0.03$. A significant improvement can be obtained by performing the same measurement under LA-phonon-assisted excitation, shown in Fig.\ref{g2}b), which allows to obtain values as low as $g^{(2)}(0) = 0.02 \pm 0.01$. This measurement was conducted at a power of \SI{10}{\micro W}, corresponding to $30\%$ of saturation power, and with a pulse duration of \SI{10}{ps}).

Similar results were obtained on sample C. Figure \ref{g2}c) represents a second-order autocorrelation measurement conducted under above-bandgap excitation on a X* transition, yielding a $g^{(2)}(0)$ value of $0.17 \pm 0.02$. This measurement was performed at a power of \SI{5}{\micro W} ($25\%$ of saturation power), and with a pulse duration of \SI{6}{ps}. In Fig.\ref{g2}d) the $g^{(2)}(0)$ value decreases to $0.04 \pm 0.02$ under LA-phonon-assisted excitation, for the same measurement performed at a power of \SI{2}{\micro W} ($4\%$ of saturation power), and with a pulse duration of \SI{16}{ps}).
The $g^{(2)}(0)$ values listed above were calculated as the ratio of the areas of the central peak over the averaged side peaks.

\textbf{\begin{figure}[h!]
\centering
\includegraphics[width=0.5\textwidth]{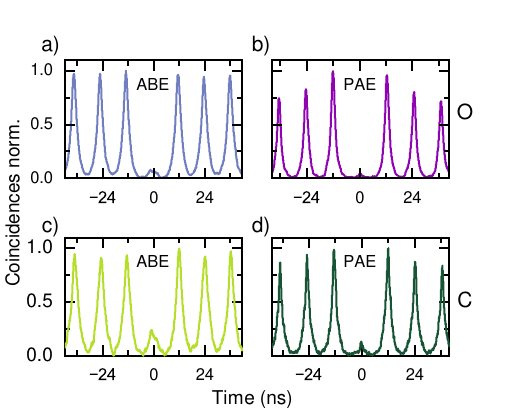}
\caption{a) Second-order autocorrelation measurement under above-bandgap excitation (ABE) on sample O, conducted at $21\%$ of saturation power, yielding a value of $g^{(2)}(0) = 0.08 \pm 0.03$. b) Measurement repeated under LA-phonon-assisted excitation (PAE) on the same X* on sample O, conducted at $30\%$ saturation power and yielding $g^{(2)}(0) = 0.02 \pm 0.01$.  c) Second-order autocorrelation measurement under ABE on sample C, conducted at $25\%$ of saturation power, yielding $g^{(2)}(0) = 0.17 \pm 0.02$. d) Measurement under PAE on the same X* on sample C, conducted at $4\%$ of saturation power, yielding $g^{(2)}(0) = 0.04 \pm 0.02$.}
\label{g2}
\end{figure}}
These values, in the range of $10^{-2}$, are consistent with those reported under the same excitation scheme on emitters with different substrates and growth techniques \cite{Zeuner2021entangled, Vyvlecka2023LA, Joos2024SUPER, phillips2024purcell, Hauser2026determ}.

In this work, we employed quasi-resonant and LA-phonon-assisted excitation to disentangle excitation-induced from intrinsic optical properties of metamorphic GaAs-based telecom quantum dots, access higher optical quality than under above-bandgap excitation, and enable coherent control. Quasi-resonant excitation narrows the trion linewidth to the setup resolution limit, the observation of Rabi-like oscillations demonstrates coherent trion control, and identifying higher excited states reveals the emitters' internal energy-level structure. Radiative lifetimes measured under these excitation schemes expose the trions’ intrinsic lifetime, and improve the single-photon purity by suppressing recapture dynamics and background emission, reaching state of the art values for telecom quantum dots. 
Beyond revealing the intrinsic characteristics of this platform, these results help establish a strategy for its further development. The demonstrated coherent optical control and high single-photon purity identify these quantum dots as promising candidates for fully resonant excitation protocols, with resonant and two-photon excitation representing natural next steps towards deterministic, high-quality single-photon generation.

Further progress towards lifetime-limited linewidths and near-unity indistinguishability will however require not only the implementation of resonant excitation schemes but also microcavity integration to enhance the photon generation rate and potentially mitigate the effects of spectral wandering. The low defect density in the active layer promoted by the growth approach makes this platform particularly attractive for deterministic integration into nanophotonic devices.

We gratefully acknowledge financial support from the Deutsche Forschungsgemeinschaft (DFG, German Research Foundation) via projects INST 95/1220-1 (MQCL) and INST 95/1654-1 (PQET), Germany’s Excellence Strategy (MCQST, EXC-2111, 390814868), the Bavarian Ministry of Economic Affairs (StMWi) via project 6GQT and the German Federal Ministry of Education and Research via the project 6G-life.

\medskip

\end{document}